\documentclass[3p,times,twocolumn]{elsarticle}

\usepackage{ecrc}

\volume{00}

\firstpage{1}

\journalname{Nuclear Physics B Proceedings Supplement}

\runauth{German Valencia}

\jid{nuphbp}

\jnltitlelogo{Nuclear Physics B Proceedings Supplement}

\usepackage{amssymb}

\usepackage[figuresright]{rotating}

\begin{document}

\begin{frontmatter}



\dochead{}

\title{Probing new physics with $\tau$-leptons at the LHC}


\author{German Valencia}

\address{Department of Physics, Iowa State University, Ames, IA 50011}

\begin{abstract}
We discuss new physics that can show up in the $\tau^+\tau^-$ production process at the LHC but not in the dimuon or the dielectron channels. We consider three different generic possibilities: a new resonance in the Drell-Yan process in the form of a non-universal $Z^\prime$; a new non-resonant contribution to $q\bar{q}\to \tau^+\tau^-$ in the form of leptoquarks; and contributions from gluon fusion due to effective lepton gluonic couplings. 
We emphasize the use of the charge asymmetry both to discover new physics and to distinguish between different possibilities
\end{abstract}


\end{frontmatter}



New physics searches in the $pp \to \tau^+\tau^-$ process are well underway in both ATLAS and CMS. From the perspective of beyond the SM physics, this channel provides a window into scenarios in which the third generation is preferred and we will discuss three such possibilities. These possibilities reinforce the need to explore  $\tau^+\tau^-$ production at the LHC regardless of limits from dimuon or dielectric channels.

A very valuable tool to measure electroweak couplings and to constrain new physics at LEP was the forward-backward asymmetry. As of now, there remain some discrepancies from the SM expectations in both $A^b_{FB}$ as measured at LEP and $A_{FB}$ in $t\bar{t}$ production as measured at the Tevatron \cite{Beringer:1900zz}.  This leads us to consider the $A^\tau_{FB}$ as well. Of course there was no measurable difference from the SM in $A^\tau_{FB}$ as measured at LEP, all the way up to CM energy of 210 GeV \cite{Abbaneo:2001ix} and this places significant constraints on new physics affecting the $\tau$-lepton that can show up at LHC, leaving mostly the high $\tau^+\tau^-$ invariant mass region to explore.

\begin{figure}
\hspace{0.5in}\includegraphics[width=2in]{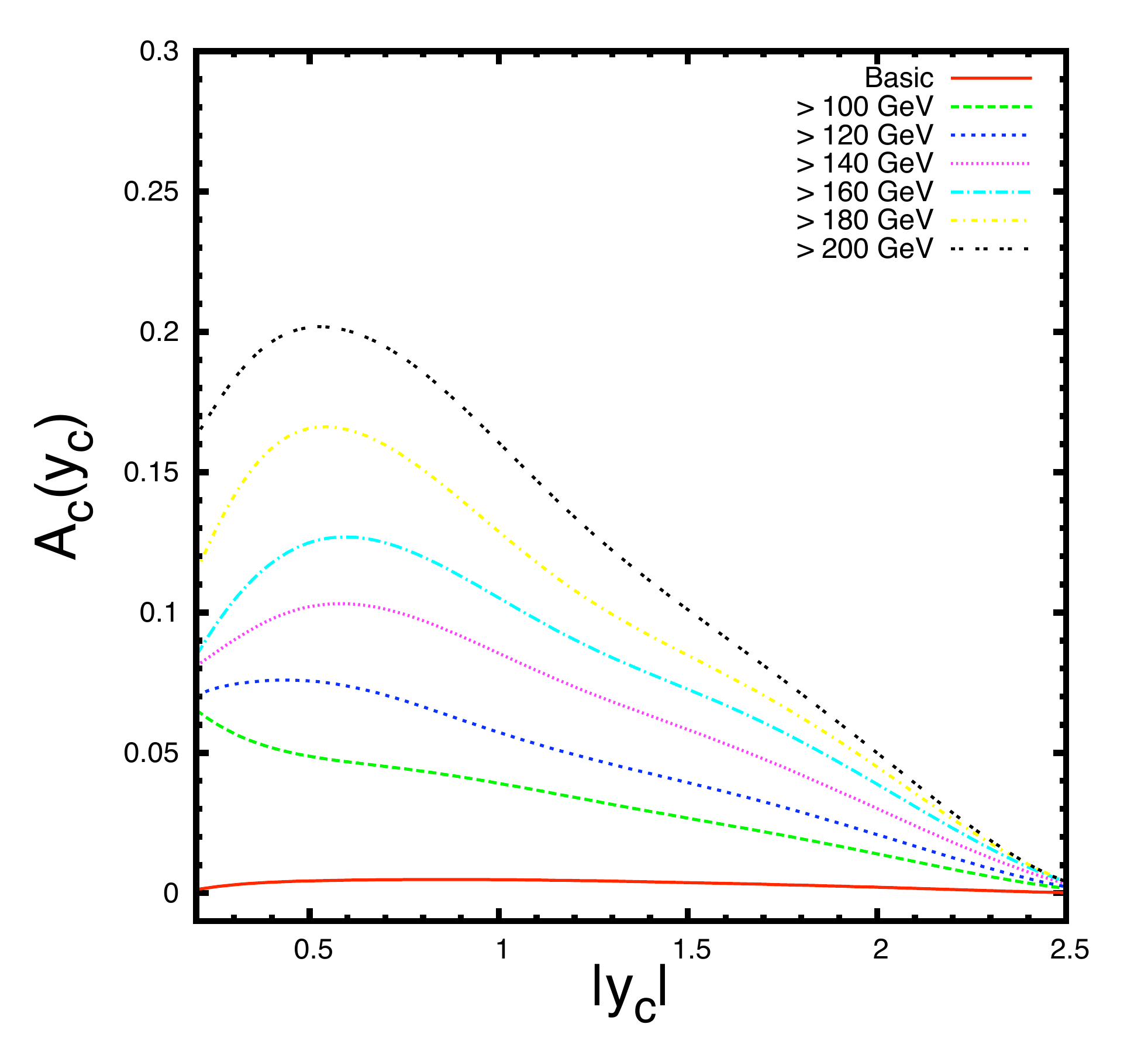}   
\caption{Integrated charge asymmetry for the standard model with basic acceptance cuts as well as different  minimum dilepton invariant mass cuts.}
\label{f:accuts}
\end{figure}

The LHC is a symmetric $pp$ collider so that one cannot define the forward backward asymmetry in the usual way. However, it is well known from corresponding studies for heavy quarks, that the information present in $A_{FB}$ can be recovered in the form of a charge asymmetry \cite{Ferrario:2009ns}. Conceptually, the simplest possibility is the reconstruction of the $q\bar{q}$ parton CM frame. If this is possible, one also knows that the direction of the quark is correlated with the direction of the boost permitting a definition of $A_{FB}$ which has already been used to measure  $A_{FB}$ for muons and electrons \cite{Chatrchyan:2012dc}, confirming SM expectations. For $\tau$-leptons it is harder to reconstruct the parton CM frame and we choose to carry our discussion in terms of the charge asymmetry. Our numerical studies show that it is better to work with an integrated charge asymmetry 
\begin{eqnarray}
{\cal A}_c (y_c) \equiv \frac{N_{\ell^+}(- y_c \leq y \leq y_c) -  N_{\ell^-}(- y_c \leq y \leq y_c)} 
{N_{\ell^+}(- y_c \leq y \leq y_c) + N_{\ell^-}(- y_c \leq y \leq y_c)}
\end{eqnarray}
The largest charge asymmetry is obtained for a value $y_c\sim 0.5$ as can be seen in Figure~\ref{f:accuts} \cite{Gupta:2011vt}
We also integrate $A_c(y_c)$ over $\tau^+\tau^-$ invariant mass from a minimum $m_{\tau\tau{\rm~min}}$ chosen at first to exclude the $Z$ region and later on to optimize the sensitivity to new physics. Figure~\ref{f:accuts} is for SM and includes basic cuts $ p_{T_\tau} > 20 {\rm~GeV}, \left|\eta_\tau \right| < 2.5, {\rm ~and~} \Delta R_{\tau\tau} > 0.4$, but we found that $y_c$ is not very sensitive to any of this. As seen in the figure, $A_c$ increases with $m_{\tau\tau{\rm~min}}$ as this has the effect of including more events with larger boosts. This comes at the price of lost statistics. An important observation is that the $\tau$-leptons at LHC are highly boosted so their decay products travel in essentially the same direction as the parent in the lab frame and this allows us to construct the asymmetry using the direction of the decay product (muon, electron or jet).

We investigate two new physics scenarios in connection with the usefulness of $A_c$. Of course we are interested in new physics scenarios that are compatible with LEP and not ruled out by measurements with muon or electron pairs so we choose the models accordingly. 

\begin{figure}[htb]
\centerline{
\includegraphics[width=1.6in]{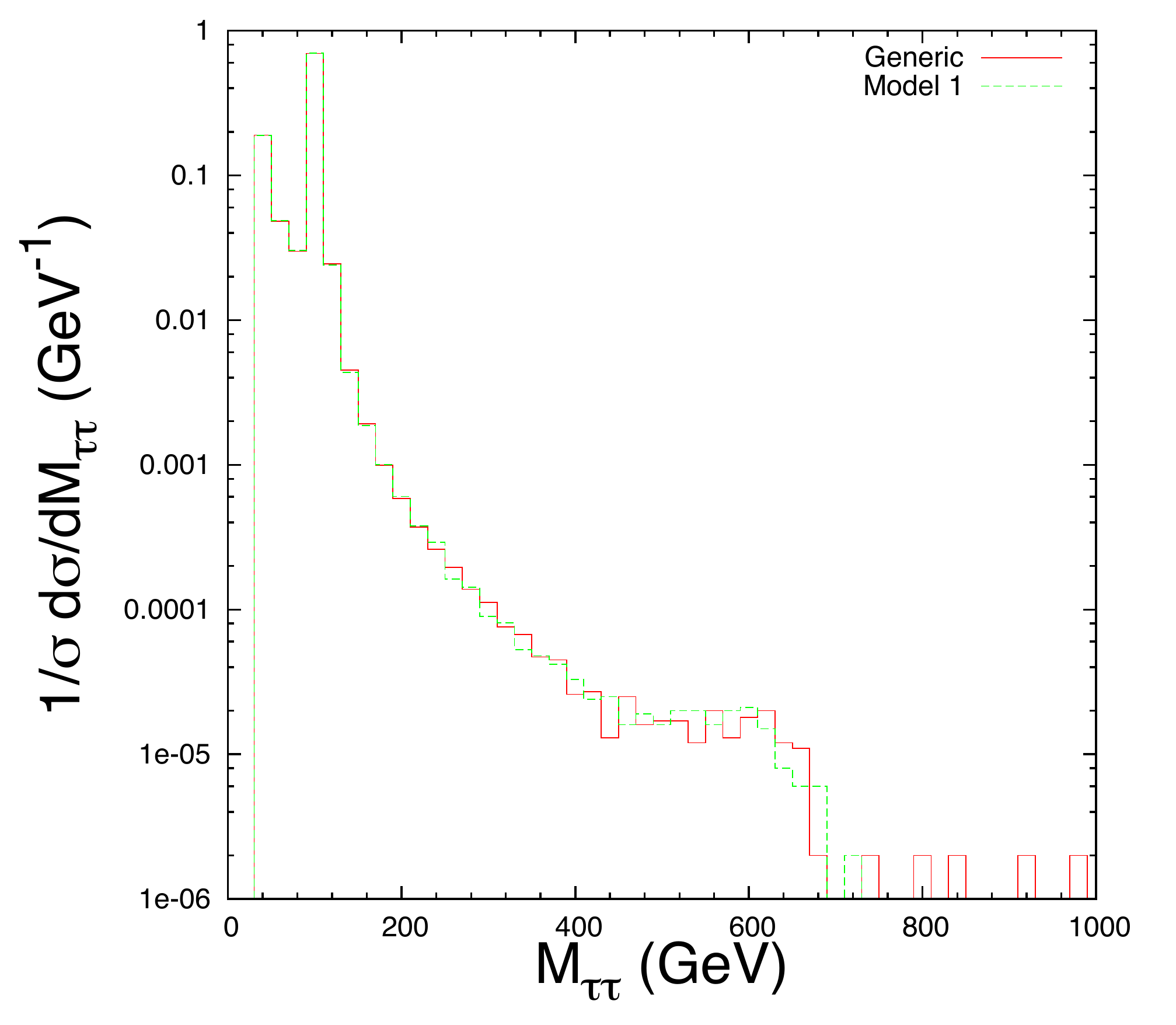}
\includegraphics[width=1.5in]{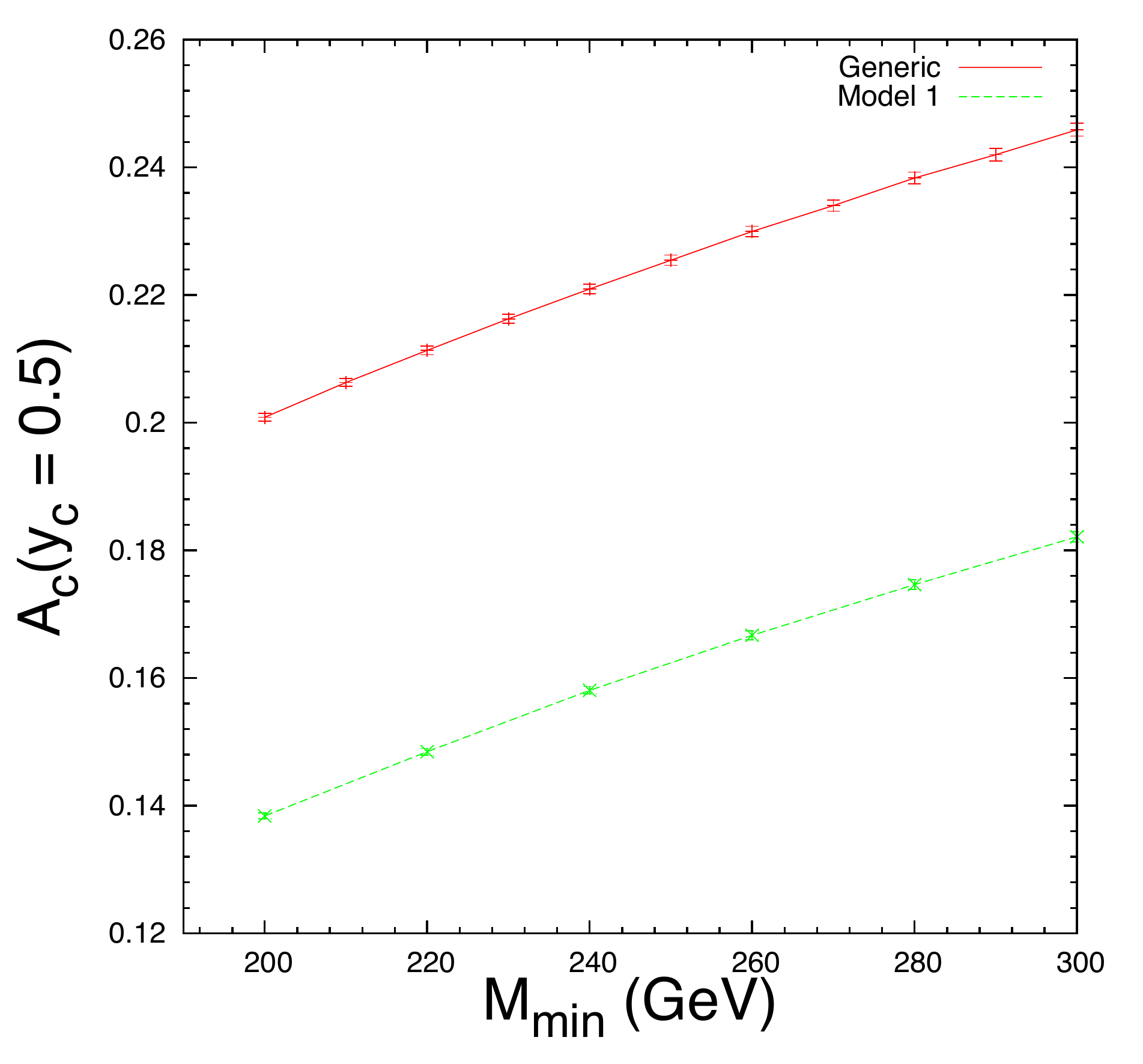}
}
\caption{\small\sf Comparison of two $Z^\prime$ models with $M_{Z^\prime}=600$~GeV in invariant mass distribution $M_{\tau\tau}$ (left) and integrated charge asymmetry (right). The model labeled `Generic'  has only non-zero couplings $c^{\tau}_R$, and $c^u_R$ such that $c^{\tau}_R \cdot c^u_R = 1/3$. The model labeled `Model 1'  is that of Ref.~\cite{He:2003qv}.}
\label{fig:Zpcompare}
\end{figure}

An example of resonant new physics of this type is a non-universal $Z^\prime$ that prefers the third generation \cite{He:2002ha,He:2003qv} (or just the $\tau$-lepton \cite{Ma:1998dp}). The main feature of such a $Z^\prime$ is that it has couplings to the third generation that are enhanced by a factor $g_R/g_L$ and couplings to the first two generations suppressed by the inverse of the same factor. In this way processes that involve only fermions from the first two generations can be suppressed below existing limits; processes involving one pair of third generation fermions, such as $e^+e^-\to \tau^+\tau^-$ or $pp\to \tau^+\tau^-$, receive corrections of electroweak strength; and processes with four third generation fermions can be significantly enhanced. Until very recently LEP2 provided the best direct bounds on this kind of resonance, but LHC has now entered the picture. CMS, for example, can exclude a $Z^\prime$ in the relevant $pp\to \tau^+\tau^-$ channel up to about 1~TeV \cite{Chatrchyan:2012hd} (although the analysis has only been done using universal $Z^\prime$ models). For comparison, the models \cite{review} analyzed by CMS are already excluded up to the 2-3~TeV range by their dimuon and dielectron analyses.  In Figure~\ref{fig:Zpcompare} \cite{Gupta:2011vt} we show the usefulness of the charge asymmetry to discriminate between two different non-universal $Z^\prime$ bosons of the same mass.  The generic couplings are of the form
\begin{eqnarray}
{\cal L}_{Z^\prime}=\frac{g}{2\cos\theta_W}\left(\bar{f}\gamma^\mu\left(c_L^fP_L+c_R^fP_R\right)f\right)Z^\prime_\mu 
\end{eqnarray}
with $c^u_R\ c^\tau_R =1/3$. 
The curve labeled `model 1' corresponds to the $Z^\prime$ of Ref.~\cite{He:2003qv} and involves  contributions from $u\bar{u}$, $d\bar{d}$ as well as strange and charm. In both cases we use a mass of 600~GeV for the $Z^\prime$. 

The events  were generated using Madgraph \cite{madgraph} and the error bars correspond to 1$\sigma$ statistical errors for 10~fb$^{-1}$ at 14~TeV.

\begin{figure}
\hspace{0.5in}\includegraphics[width=2.in]{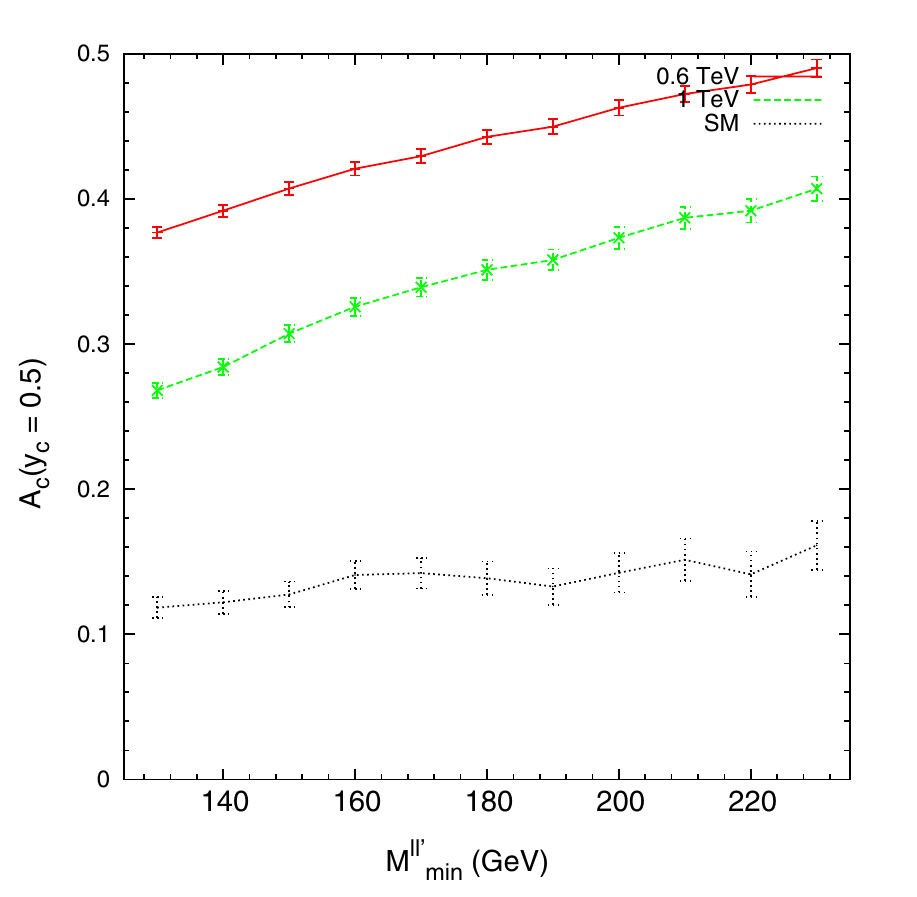}   
\caption{Charge asymmetry in the $\tau_e\tau_\mu$ mode  for the SM and the LQ-2 model with two different lepto-quark masses as a function of a minimal $M_{\ell\ell^\prime}$ cut.}
\label{lq-dilepton}
\end{figure}

As an example of non-resonant new physics that would affect $\tau$-pair production in the Drell-Yan process we next consider the exchange of leptoquarks (LQ). The generic couplings of vector LQ are given by  \cite{leptoquarks}
\begin{eqnarray}
&&{\cal L}_{LQ} = {\cal L}_{SM} \\\nonumber
&&+ \lambda^{(R)}_{V_0}\cdot \overline{d} \gamma^{\mu} P_R e
\cdot V_{0\mu}^{R\dagger} +
\lambda^{(R)}_{\tilde V_0}\cdot \overline{u} \gamma^{\mu} P_R e \cdot \tilde{V}_{0\mu}^{\dagger} \\ \nonumber
&&+ \lambda^{(R)}_{V_{1/2}}\cdot \overline{d^c} \gamma^{\mu} P_L \ell
\cdot {V}_{1/2\mu}^{R\dagger} +
\lambda^{(R)}_{\tilde V_{1/2}}\cdot \overline{u^c} \gamma^{\mu} P_L \ell \cdot
\tilde{V}_{1/2\mu}^{\dagger} \\ \nonumber
&&+
\lambda^{(L)}_{V_0}\cdot \overline{q} \gamma^{\mu} P_L \ell \cdot
V_{0\mu}^{L\dagger} +
\lambda^{(L)}_{V_{1/2}}\cdot \overline{q^c} \gamma^{\mu} P_R e \cdot
V_{1/2\mu}^{L\dagger} \\ \nonumber
&&+ \lambda^{(L)}_{V_1}\cdot \overline{q} \gamma^{\mu}  P_L
{V}_{1\mu}^{\dagger} \ell + h.c.
\end{eqnarray}
LQ that would affect primarily this process, for example, are Pati-Salam vector LQ with a strong coupling $\lambda^{(R)}_{V_0}=\lambda^{(L)}_{V_0} = g_s/\sqrt{2}$ and with quantum numbers that couple the first generation quarks to the third generation fermions \cite{Valencia:1994cj}. These would contribute to $pp\to \tau^+\tau^-$ via a t-channel $d\bar{d}\to \tau^+\tau^-$ diagram. Since the protons have more up quarks than down quarks, we consider instead a variation of this model  in which the LQ has charge $5/3$ and contributes via a t-channel $u\bar{u}\to \tau^+\tau^-$ diagram, $\lambda^{(R)}_{\tilde{V}_0} = g_s/\sqrt{2}$, we call this model LQ2. In Figure~\ref{lq-dilepton}
we show how the charge asymmetry could easily differentiate between the SM and LQ2 for a leptoquark mass 1~TeV which would be hard to detect in the lepton-pair invariant mass distribution. 
The plot is for 14~TeV with cuts $p_T > 6~{\rm~GeV}, |\eta| < 2.5 {\rm ~and~} \Delta R_{ik} < 0.4, \, i, k = \ell, j$. We show the $\tau_e\tau_\mu$ channel which has less background. The $\tau_e\tau_e$ and $\tau_\mu\tau_\mu$ modes can also be used with the additional requirement of a minimum missing $E_T$ (we used 10~GeV) that removes the direct Drell-Yan production of dimuons or dielectrons. In all cases the charge asymmetry is constructed using the direction of the decay lepton and the error bars correspond to 1$\sigma$ statistical errors for 10~fb$^{-1}$ at 14~TeV \cite{Gupta:2011vt}. 

\begin{figure}
\hspace{0.1in}\includegraphics[width=3.0in]{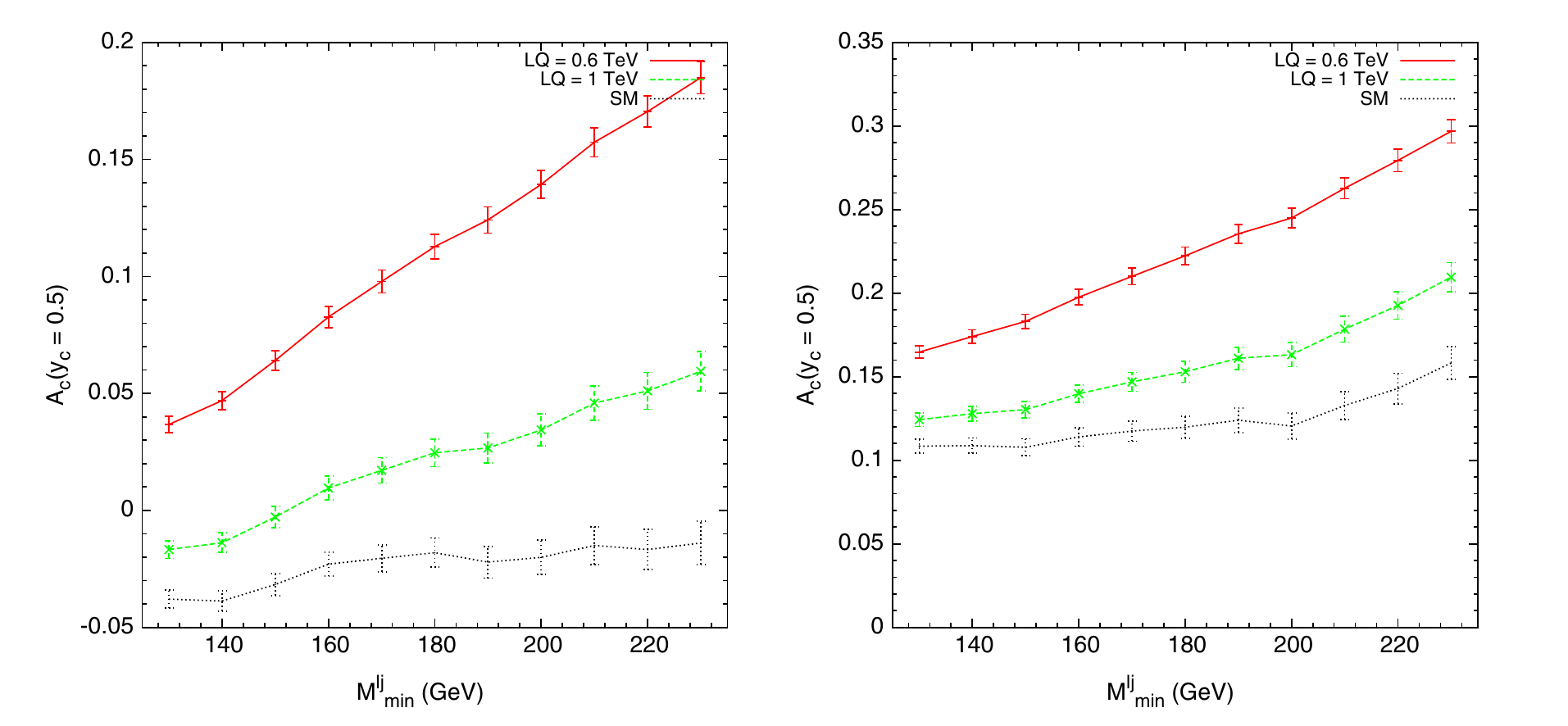}  
\caption{Charge asymmetry $A_{lj}$ with a missing $E_T$ requirement. The results for $\ell^+ j$ and $\ell^- j$ are shown respectively in the left-side and right-side plots}
\label{lp_jet}
\end{figure}

The analysis can also be carried out in the $\tau_e\tau_h$ and $\tau_\mu\tau_h$ modes as we have shown in Ref.~\cite{Gupta:2011vt}. For $\tau_h$ we included only the one pion and one rho modes, and used a 0.3\% probability of a QCD jet in $Wj$ events faking a $\tau$. The Figure~\ref{lp_jet} uses the same cuts described above and includes background from $W$ pairs, $Z$ pairs, and $Wj$ events. It shows that 
it is possible to discriminate between LQ2 and the SM using these decay channels. The two plots are different because the $W^+j$ background is larger than the $W^-j$ background and they also have different charge asymmetries (-6.3\% and 11\% with our set of cuts).

\begin{figure}
\hspace{0.1in}\includegraphics[width=3.2in]{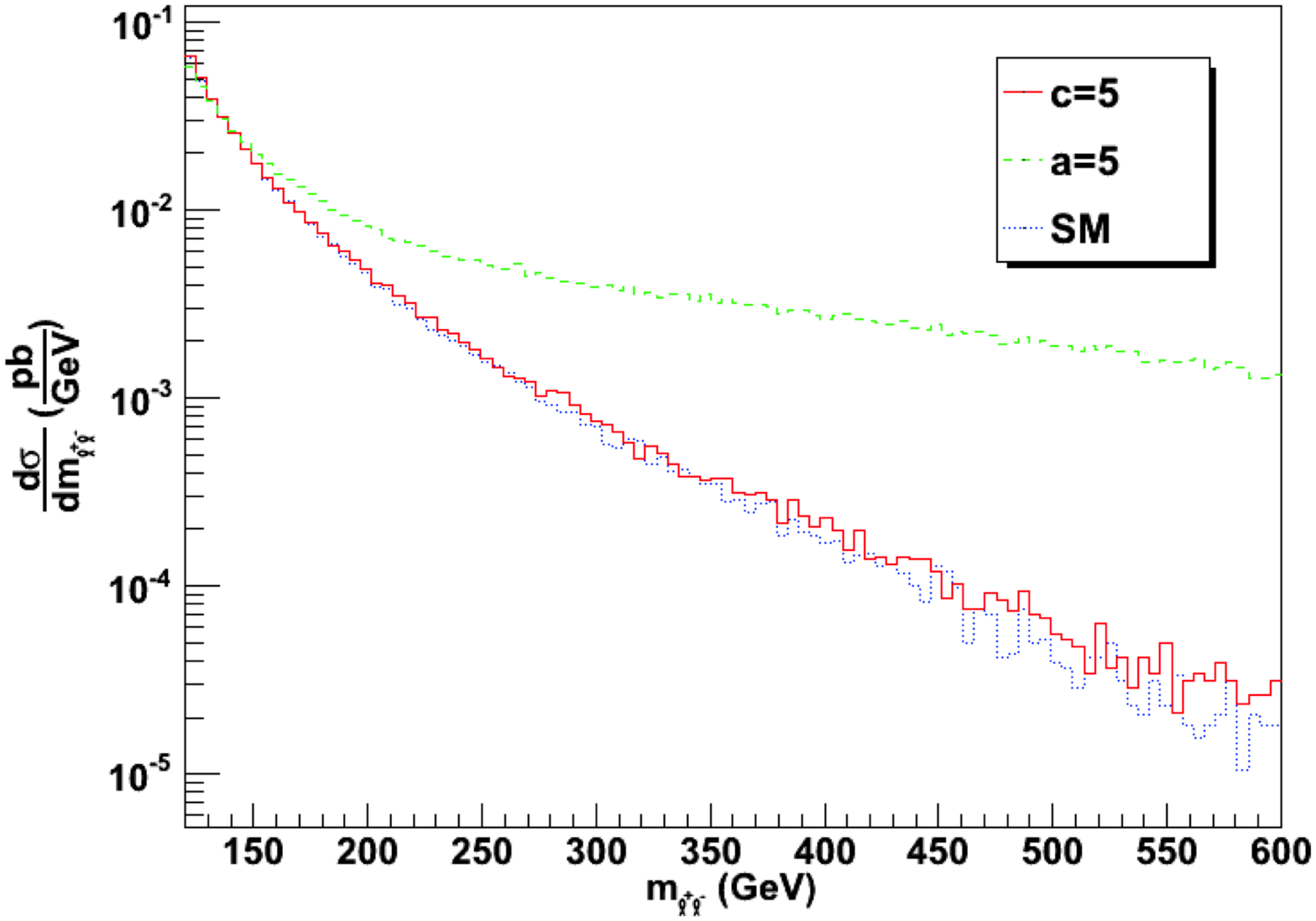} 
\hspace{0.2in}\includegraphics[width=3.2in]{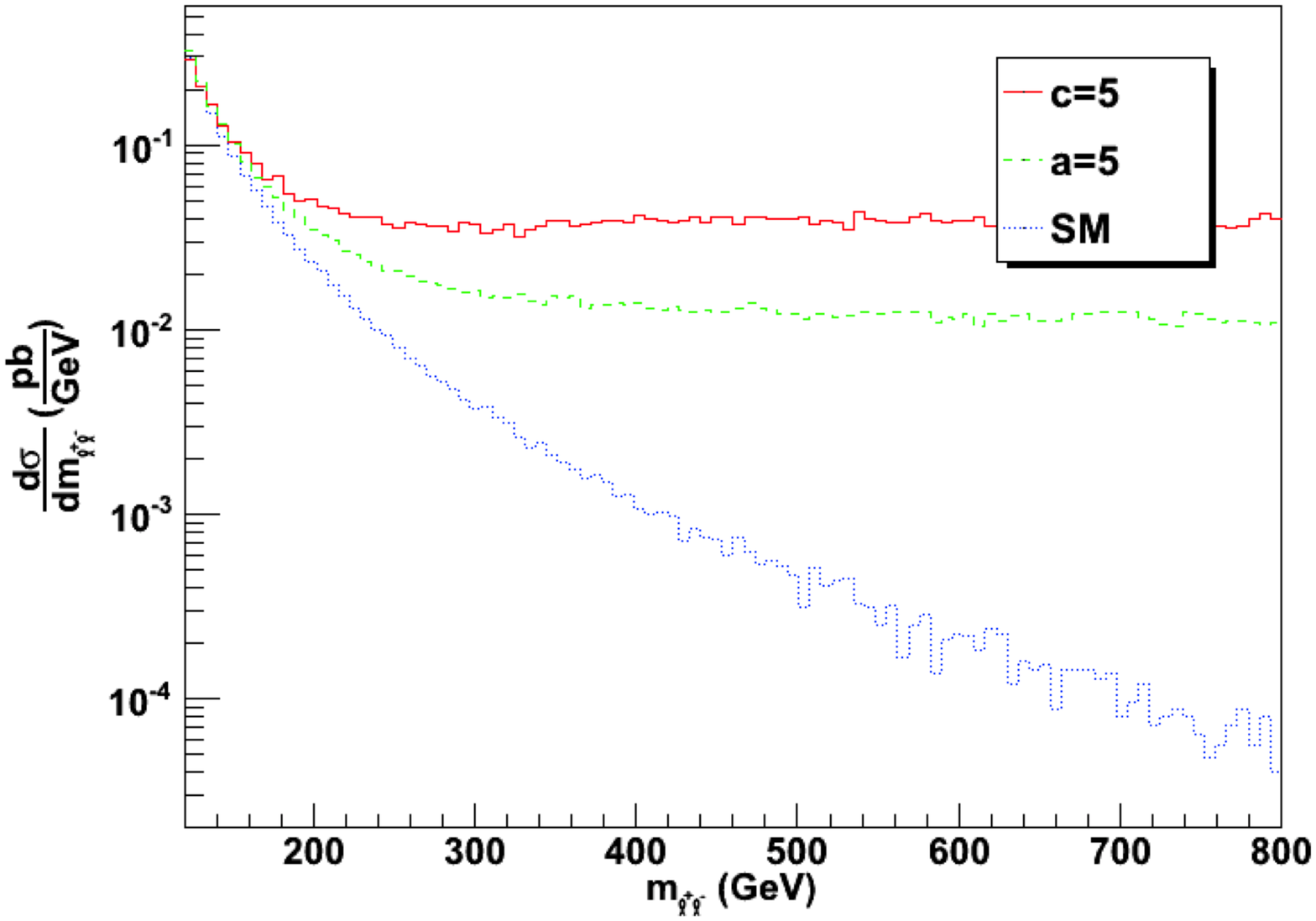}  
\caption{$d\sigma/dm_{\ell\ell}$ for the SM; the SM plus new physics in the gluon fusion process ($c=5$); and the SM plus new physics in the $u\bar{u}$ annihilation process ($a=5$). The scale of new physics is taken to be $\Lambda =1$~TeV. The top figure corresponds to $p\bar{p}\to \ell^+\ell^-$ at the Tevatron, and the bottom figure to $pp\to \ell^+\ell^-$ at the LHC.}
\label{gl-tev}
\end{figure}

Finally we turn our attention to the possibility of new physics contributing to lepton pair production via gluon fusion. This would be a very interesting possibility since the LHC is a `gluon collider' and is not as exotic as it first seems. An example of new physics that connects gluons to leptons is a new heavy neutral Higgs which has already been studied at LHC \cite{Aad:2011rv,Chatrchyan:2011nx}. We are more interested in a `lepton gluonic coupling': an effective coupling between gluons and leptons away from a new resonance. The natural formalism to describe this scenario is the effective Lagrangian \cite{Buchmuller:1985jz} in which one writes corrections to the SM in the form of higher dimension operators suppressed by the scale of new physics. Schematically,
\begin{eqnarray}
{\cal L} = {\cal L}_{SM} + \frac{1}{\Lambda^2}(\sum_i{\cal O}_i^{d=6})+ \frac{1}{\Lambda^4}(\sum_i{\cal O}_i^{d=8})+\cdots
\end{eqnarray}
For processes at an energy scale $E$, the higher dimension operators contribute amplitudes suppressed by increasing powers of $E/\Lambda$ and for this reason one limits these studies to the lowest dimension, usually six. At the LHC, however, the large parton luminosity for gluon-gluon interactions distorts this power counting, enhancing gluon fusion initiated processes. This makes the `lepton gluonic couplings' that first appear at dimension eight possibly competitive with other dimension six operators. There are two such operators (and their hermitian conjugates):
\begin{eqnarray}
{\cal L} = \frac{g_s^2}{\Lambda^4}\left(c \ G^{A\ \mu\nu}G^A_{\mu\nu} \bar \ell_L \ell_R \phi  +\tilde{c} \ G^{A\ \mu\nu} \tilde G^A_{\mu\nu} \bar \ell_L \ell_R \phi \right).
\end{eqnarray}
Ignoring CP violating phases, the two operators affect the parton cross-sections in identical manner:
\begin{eqnarray}
\frac{d\hat{\sigma}(gg\to \ell^+\ell^-)}{d\hat{t}}&=&\left(|c|^2+|\tilde{c}|^2\right)\frac{v^2 g_s^4}{\Lambda^8}\frac{\hat{s}}{32\pi}, \nonumber \\
\hat{\sigma}(gg\to \ell^+\ell^-)&=&\left(|c|^2+|\tilde{c}|^2\right)\frac{v^2 g_s^4}{\Lambda^8}\frac{\hat{s}^2}{32\pi}.
\end{eqnarray}
The lepton flavor structure of the operators can be any, including lepton flavor violating, but here we concentrate on the case of $\tau$-flavor. In  the top Figure~\ref{gl-tev} we compare the effect of the lepton gluonic coupling operator with that of a dimension six operator of the form $a g^2/\Lambda^2 \bar{u}u\bar\ell \ell$ (chosen so that it doesn't interfere with the SM) at the Tevatron \cite{Potter:2012yv}. 
The figure confirms our expectation of dominance by the dimension six operator. At the LHC, however, the situation is much different, due to the enhancement of gluon fusion, and we illustrate this in the bottom Figure~\ref{gl-tev} \cite{Potter:2012yv}. 

One example of a model that could induce these lepton gluonic couplings involves a heavy scalar with couplings to fermions as in a 2HDM with large $\tan\beta$. With a heavy fourth generation one could get $c\sim m_\ell \tan^2\beta/v$ with $\Lambda^4 \sim (4\pi v)^2M_S^2$. Another possibility would be a model with a vector LQ of mass $M_X$ coupling heavy quarks to tau leptons. In this case one would find  $c\sim\pi\alpha_s$ and $\Lambda^4 \sim (4\pi v)^2M_X^2$.

Our numerical simulation with the aid of Madgraph \cite{madgraph} indicates that the LHC has a $3\sigma$ statistical sensitivity to $c\gtrsim 4.3$ for $\Lambda =2$~TeV and 10~fb$^{-1}$ at 14~TeV. This compares to the Tevatron's  $3\sigma$ statistical sensitivity to $c\gtrsim 75$, LEP2's $c\gtrsim 80$ and the partial wave unitarity constraint $c\lesssim 80$.

To summarize, we have investigated three different scenarios that illustrate new physics that can be constrained at LHC by studying $\tau$-lepton pair final states:
\begin{enumerate}[(i)]
\item As an example of a new resonance in the Drell-Yan process we considered a non-universal $Z^\prime$. We showed how the charge asymmetry can distinguish between different possibilities of new resonances with the same mass.
\item As an example of non-resonant contributions to Drell-Yan, we considered vector leptoquarks. In this case we saw that the charge asymmetry can signal the presence of new physics even when it is not visible as a bump in an invariant mass distribution.
\item Finally we considered the case of lepton gluonic couplings which first occur at dimension eight and which can be tested for the first time at LHC.
\item It is worth emphasizing that the lepton gluonic couplings can occur in any dilepton channel, including those that violate lepton flavor, and that all of them should be investigated.
\end{enumerate}

{\bf Acknowledgements:} 
This talk was based on work done with Sudhir Gupta and Hugh Potter and supported in part by the DOE under contract number
DE-FG02-01ER41155.




\nocite{*}
\bibliographystyle{elsarticle-num}
\bibliography{martin}

\begin{thebibliography}{00}


\bibitem{Beringer:1900zz} 
  J.~Beringer {\it et al.}  [Particle Data Group Collaboration],
  Phys.\ Rev.\ D {\bf 86}, 010001 (2012).

\bibitem{Abbaneo:2001ix} 
  D.~Abbaneo {\it et al.}  [ALEPH and DELPHI and L3 and OPAL and LEP Electroweak Working Group and SLD Heavy Flavor and Electroweak Groups Collaborations],
  hep-ex/0112021.
 
\bibitem{Ferrario:2009ns}
  P.~Ferrario and G.~Rodrigo,
  J.\ Phys.\ Conf.\ Ser.\  {\bf 171}, 012091 (2009)
  [arXiv:0907.0096 [hep-ph]].

\bibitem{Chatrchyan:2012dc} 
  S.~Chatrchyan {\it et al.}  [CMS Collaboration],
  arXiv:1207.3973 [hep-ex].
  
\bibitem{Gupta:2011vt} 
  S.~K.~Gupta and G.~Valencia,
  Phys.\ Rev.\ D {\bf 84}, 036009 (2011)
  [arXiv:1102.0741 [hep-ph]].

\bibitem{He:2002ha}
  X.~G.~He and G.~Valencia,
  Phys.\ Rev.\  D {\bf 66}, 013004 (2002)
  [Erratum-ibid.\  D {\bf 66}, 079901 (2002)]
  [arXiv:hep-ph/0203036];.

\bibitem{He:2003qv}
  X.~G.~He and G.~Valencia,
  Phys.\ Rev.\  D {\bf 68}, 033011 (2003)
  [arXiv:hep-ph/0304215].

\bibitem{Ma:1998dp}
  E.~Ma, D.~P.~Roy,
  Phys.\ Rev.\  {\bf D58}, 095005 (1998).
  [hep-ph/9806210].
  
\bibitem{Chatrchyan:2012hd} 
  S.~Chatrchyan {\it et al.}  [CMS Collaboration],
  Phys.\ Lett.\ B {\bf 716}, 82 (2012)
  [arXiv:1206.1725 [hep-ex]].
  
\bibitem{review} For a recent review see
  P.~Langacker,
  Rev.\ Mod.\ Phys.\  {\bf 81}, 1199 (2008)
  [arXiv:0801.1345 [hep-ph]].
  
  
 \bibitem{madgraph}
  T.~Stelzer and W.~F.~Long,
  Comput.\ Phys.\ Commun.\  {\bf 81}, 357 (1994)
  [arXiv:hep-ph/9401258];
  J.~Alwall {\it et al.},
  JHEP {\bf 0709}, 028 (2007)
  [arXiv:0706.2334 [hep-ph]];
  J.~Alwall, M.~Herquet, F.~Maltoni, O.~Mattelaer, T.~Stelzer,
  JHEP {\bf 1106}, 128 (2011).
  [arXiv:1106.0522 [hep-ph]]. 
  
\bibitem{leptoquarks}
  J.~C.~Pati and A.~Salam,
  Phys.\ Rev.\  D {\bf 10}, 275 (1974)
  [Erratum-ibid.\  D {\bf 11}, 703 (1975)]; 
  S.~Davidson, D.~C.~Bailey and B.~A.~Campbell,
  Z.\ Phys.\  C {\bf 61}, 613 (1994)
  [arXiv:hep-ph/9309310]; 
  M.~Hirsch, H.~V.~Klapdor-Kleingrothaus and S.~G.~Kovalenko,
  Phys.\ Lett.\  B {\bf 378}, 17 (1996)
  [arXiv:hep-ph/9602305].
  
\bibitem{Valencia:1994cj}
  G.~Valencia and S.~Willenbrock,
  Phys.\ Rev.\  D {\bf 50}, 6843 (1994)
  [arXiv:hep-ph/9409201].
 
\bibitem{Aad:2011rv} 
  G.~Aad {\it et al.}  [ATLAS Collaboration],
  Phys.\ Lett.\ B {\bf 705}, 174 (2011)
  [arXiv:1107.5003 [hep-ex]].
  
\bibitem{Chatrchyan:2011nx} 
  S.~Chatrchyan {\it et al.}  [CMS Collaboration],
  Phys.\ Rev.\ Lett.\  {\bf 106}, 231801 (2011)
  [arXiv:1104.1619 [hep-ex]].

\bibitem{Buchmuller:1985jz} 
  W.~Buchmuller and D.~Wyler,
  Nucl.\ Phys.\ B {\bf 268}, 621 (1986).
  
\bibitem{Potter:2012yv} 
  H.~Potter and G.~Valencia,
  Phys.\ Lett.\ B {\bf 713}, 95 (2012)
  [arXiv:1202.1780 [hep-ph]].


\end{thebibliography}



\end{document}